# Energy-efficient superparamagnetic Ising machine and its application to traveling salesman problems

Jia Si[1,2], Shuhan Yang[1], Yunuo Cen[1], Jiaer Chen[1], Zhaoyang Yao[1], Dong-Jun Kim[1], Kaiming Cai[1], Jerald Yoo[1], Xuanyao Fong[1] and Hyunsoo Yang[1]*

[1]*Department of Electrical and Computer Engineering, National University of Singapore, Singapore.*

[2]*Key Laboratory for the Physics and Chemistry of Nanodevices and Center for Carbon-based Electronics, School of Electronics, Peking University, Beijing, China.*

*e-mail: eleyang@nus.edu.sg

**The growth of artificial intelligence and IoT has created a significant computational load for solving non-deterministic polynomial-time (NP)-hard problems, which are difficult to solve using conventional computers. The Ising computer, based on the Ising model and annealing process, has been highly sought for finding approximate solutions to NP-hard problems by observing the convergence of dynamic spin states. However, it faces several challenges, including high power consumption due to artificial spins and randomness emulated by complex circuits, as well as low scalability caused by the rapidly growing connectivity when considering large-scale problems. Here, we present an experimental Ising annealing computer based on superparamagnetic tunnel junctions (SMTJs) with all-to-all connections, which successfully solves a 70-city travelling salesman problem (4761-node Ising problem). By taking advantage of the intrinsic randomness of SMTJs, implementing a proper global annealing scheme, and using an efficient algorithm, our SMTJ-based Ising annealer shows superior performance in terms of power consumption and energy efficiency compared to other Ising schemes. Additionally, our approach provides a promising way to solve complex problems with limited hardware resources. Moreover, we propose a crossbar array architecture for scalable integration using conventional magnetic random access memories. Our results demonstrate that the**

**SMTJ-based Ising annealing computer with high energy efficiency, speed, and scalability is a strong candidate for future unconventional computing schemes.**

The demands for future data-intensive and energy-efficient computing tasks overwhelm the computational power of conventional von Neumann architectures[1]. For example, NP-hard problems are often encountered in combinatorial optimizations[2], resource allocation[3], cryptography[4], finance[5], image processing[6], tour planning[7], and job sequencing[8], and their computational time and hardware resources increase exponentially with the problem size, which makes them very difficult or impossible to be solved by conventional computers in a finite time. These problems can be mapped to the Ising model, a mathematical model to characterize interactions between magnetic spins[9]. The dynamics of the model is algorithm-based, i.e. by constructing a proper coupling matrix and allowing the system to evolve utilizing an intrinsic convergence property of the Ising model, the ground state could be obtained as a solution to the corresponding problems. However, as the system might be trapped in many local minima, the annealing process has usually been adopted in Ising computers to address such limitations. It is commonly agreed that adding fluctuations prevents the Ising computer from being stuck at the local minima.

Efficient algorithms and hardware systems for finding an optimal or near-optimal solution of an Ising model at a fast speed and low power have been sought. Adiabatic quantum computing (AQC)[10,11] and quantum computing[12–15] based on superconducting qubits are capable of converging the Ising model by tunnelling out of local minima to the global minima. A 100-node Maxcut problem was solved using a quantum computer of 2048 spins with huge power consumption[16]. Besides the high cost and complexity of cryogenic temperature, this proof-of-concept system was limited by the sparse connections only between the nearest neighbors, which leads to sub-optimal outcomes[17]. Simulated annealing based on CMOS implementations was exploited for parallel Ising computing, including central processing units (CPU)[18,19], graphics processing units (GPU)[20], and field-programmable gate array (FPGA)[21,22]. These hardware have reported as large as 16,384 spins, however, it requires huge hardware resources for generating random numbers to introduce stochasticity to escape from the local minima[4,18,23,24]. Coherent Ising machine (CIM) is an optical scheme with competitive energy efficiency. However, it requires a long fibre ring cavity and relies on external FPGA for implementing coupling[25,26]. The temporal multiplexing process is also time-consuming and hard to expand to large systems. Recently, experiments and simulation works have investigated various devices to emulate the behavior of Ising spins by taking advantage of their intrinsic physics. An 8-spin asynchronous probabilistic computer based on superparamagnetic tunnel junctions for solving integer factorization tasks of values up to 945 was demonstrated[4]. SPICE simulations of 16-city TSP using simulated annealing method were presented[27]. Other works such as 8-spin phase-transition nano-oscillators[28], multiferroic oxide devices with a high thermal stability[29], and magnetoresistive random access memory (MRAM)[30,31] have also conceptually proved that spin-based devices are suitable for representing Ising units. However, these works have encountered challenges in either partially-connected Ising spins or small scalability which limit the Ising computer from solving practical problems. TSP discussed in this paper is a well-known problem which is much beyond the limitation of locally connected Ising models. Other combinatorial optimization problems, such as knapsack problems, colouring problems, and number partitioning, need all-to-all connection to satisfy specific constraints[9]. In practice, an additional graph embedding process is often required when mapping to 2-dimensional CMOS circuitry which only considered the coupling between adjacent spins[32–34]. Since the embedding increases the required number of auxiliary spins and causes spin connections to change, the annealing accuracy is degraded significantly, especially when the problem size is large. This means that supporting a fully connected Ising model is highly recommended for dealing with a wide range of problems. Another problem is the rapidly increasing connectivity when considering large-scale systems, which usually results in huge energy consumption and latency. Since the number of spins that a particular annealing processor can handle limit the scale of the problem that can be solved, how to solve complex problems with limited hardware in an energy-efficient way has also drawn significant attention.

In this work, we experimentally report a scalable Ising computer based on 80 SMTJs with all-to-all connections and successfully solve the 4761-node TSP problem. The intrinsic stochasticity in SMTJ enables ultra-fast and low-power Ising annealing without using extra resources for random number generation and Metropolis determining process[7]. By combining global annealing with intrinsic annealing in SMTJ, the convergence of the Ising problem is guaranteed especially in large-scale Ising problems. With an all-to-all connection among Ising spins, the combinatorial optimization of 9-city TSP is solved with the optimal solution. We further develop the algorithm for CTSP with no extra auxiliary Ising bits both in algorithm and hardware, indicating the superiority and flexibility of this Ising computer. Furthermore, we propose an optimization strategy based on GP and CTSP and experimentally solved a 70-city TSP, which typically needs 4761 nodes, on our 80-node Ising computer with a near-optimal solution. The system can obtain the lowest power consumption of 0.64 mW as well as high energy efficiency of 39 solutions per second per watt among state-of-art Ising annealers. We have experimentally demonstrated that large-scale Ising problems can be solved by small-scale hardware in an energy-efficient way.

**Results**

**SMTJ-based artificial Ising spin**

Various NP-hard problems can be solved by constructing corresponding Ising models and observing the ground states during evolution processes. Fig. 1a shows an all-to-all connected Ising model, whose Ising Hamiltonian can be written as

$$H = -\sum_{i,j}^{N} J_{i,j} s_i s_j - \sum_{i}^{N} h_i s_i, \tag{1}$$

where $H$ is the total energy of the system, $N$ is the total number of spins, $s_i$ is the $i$-th spin with one of two states; "+1" (Ising spin up) or "−1" (Ising spin down), $J_{i,j}$ is the coefficient of coupling between the $i$-th and the $j$-th spins, and $h_i$ is the external field of the $i$-th spin. For a fixed configuration of other spins than $s_k$, the probability of $s_k$ staying in the down-state is given by

$$p_\downarrow = \frac{1}{1+e^{-2\Lambda}}. \tag{2}$$

where $\Lambda = \frac{\partial H}{\partial s_k}$ (see Supplementary Note 1).

One natural implementation of this Ising spin is based on a stochastic nanomagnet. The inset of Fig. 1b shows the sketch of an SMTJ, consisting of a tunnelling barrier layer sandwiched by a reference layer and a free layer (see Methods section). Because of the small device diameter (~ 50 nm) of the device, the energy barrier of the magnetic free layer between the anti-parallel (AP) and parallel (P) states is low that the retention time of either state is in the range of μs to ms, similarly reported in previous studies[4,35]. The SMTJ resistance, measured as a function of time in Fig. 1c, shows preferred AP states at high currents and P states at low currents. When the current ($I$) is ~4 μA, SMTJ shows an equal chance of AP and P states. The probability of the AP state under different input currents over 0.1 s is fitted in Fig. 1b by a sigmoid function:

$$p_{\_AP} = \frac{1}{1+e^{-a \times (I-b)}} \tag{3}$$

where a = 4.67 and b = 3.9 μA. In order to emulate Ising spin $s_k$ with our SMTJ device, we only need to make the probability of the down-state of $s_k$ to be equal to that for the AP state of SMTJ, namely $p_{\_AP} = p_\downarrow$, with two calibration coefficients. Thus, we can derive the form of the current $I_{\_k}$ injected to SMTJ as (see Supplementary Note 1):

$$I_{\_k} = \frac{2\Lambda}{a} + b = \frac{c}{a}\left(\sum_j 2J_{kj}s_j + h_k\right) + b \tag{4}$$

where $c = 1/kT$ is the effective inverse temperature which can be conducted for global annealing.

**Intrinsic annealing in SMTJs-based Ising computer**

By integrating 80 SMTJs with a peripheral circuit and a microcontroller unit (MCU), we build an 80-node Ising computer (see Supplementary Note 2). Each Ising spin in Eq. (1) is emulated by an SMTJ with intrinsic randomness, where P (AP) state represents spin-up (down). Fig. 1d shows the photograph of the printed circuit board (PCB) and the diagram of the system (see Methods section). The system contains 8 processing elements (PEs); each PE has 10 SMTJ computing units. Each SMTJ computing unit includes a transistor and a resistor to adjust the state of SMTJ into stochastic. During the computing process, an MCU examines the states of all SMTJs by reading the output of comparator arrays through multiplexers and generates new input voltages for digital-to-analogue converters (DACs) according to the updating rule in Eq. (4) (see Supplementary Note 3 for calibration of 80 SMTJ computing units).

During the evolution process, an Ising solver could be easily trapped in a local minimum state. To avoid this non-optimal solution, annealing algorithms such as simulated annealing (SA) or quantum annealing (QA) were developed. The general idea of SA is to make the system evolve from a high temperature to a low temperature gradually[7]. The convergence and relaxation of simulated annealing can be mathematically provable[36]. During each iteration, a random number is generated for stochasticity and introduced to determine whether the result in this iteration should be accepted or not. In QA, quantum fluctuations cause quantum tunnelling between states[17]. In both SA and QA, stochasticity needs to be introduced into the annealing process. In contrast, our Ising system utilizes the intrinsic stochastic behaviours of SMTJ to perform the Metropolis process of standard SA in hardware, which greatly saves the solution time and hardware resources for generating randomness (see Supplementary Note 4).

Besides, our Ising computer has an all-to-all connection which has wider application scenarios, as well as a better capability of escaping from local minima.

**Ising Mapping of N-city TSP and CTSP**

We have applied our Ising computer to the TSP problem, one of the combinatorial optimization problems, which applies to various sectors, such as vehicle routing, logistics, planning, and scheduling. The goal is to find the shortest route that visits all listed cities once and only once given distances between the cities in the list. In order to solve this problem, we first map $N$-city-TSP to an $N^2$-spin Ising model, or $(N-1)^2$-spin model assuming a fixed starting city. Fig. 2a shows the coordinates of 9 cities and Fig. 2b shows the 81-spin Ising model, whose rows indicate the name of the city and columns indicate the visiting order. We define the binary spin, $s$, as $s_{i,j} = 1$ if city $i$ is visited as $j$-th city or $s_{i,j} = -1$ otherwise. The total Hamiltonian of TSP is expressed by[9]

$$H_{TSP} = \sum_i (\sum_j s_{ij} + (N-2))^2 + \sum_j (\sum_i s_{ij} + (N-2))^2 + w \sum_j \sum_{i,i'} d_{i,i'} (\frac{s_{i,j}+1}{2})(\frac{s_{i',j+1}+1}{2}) \quad (5)$$

where the first term is a constraint that represents only one city is visited at the $j$-th visit, and the second term is another constraint that represents one city is visited only one time. $w$ is a constant small enough ($0 < w < 1$) not to violate the two constraints of the TSP cycle. $d_{i,i'}$ is the distance between city $i$ and city $j$. According to Eq. (1) and (5), coupling matrix $J$ of 81 spins could be obtained, as shown in Fig. 2c (see Supplementary Note 5). It shows that spins in the same row or column have strong coupling, as indicated by the first two terms in Eq. (5).

We define CTSP as the visiting orders of some cities are enforced during the travelling. This is quite useful in real-life scenarios. For example, a delivery man collects food and drinks at shop $A$ and must deliver hot drinks to $B$ first even though the total cost is higher than optimal. We propose an algorithm for solving CTSP by adding negative "distance" to the Hamiltonian. For example, suppose city A and city B are required to be connected in the CTSP as city 2 and city 7 shown in Fig. 2d, then we add the term

$$-\theta H_c = -\theta(\sum_j s_{A,j}s_{B,j+1} + s_{B,j}s_{A,j+1}). \quad (6)$$

such that the energy of a path, where city A and city B are connected, is always lowered by $\theta$. When $\theta$ is sufficiently large, the optimal path must have city 2 and city 7 connected. Thus, the total Hamiltonian of the CTSP is expressed by

$$H_{CTSP} = H_{TSP} - \theta H_c. \quad (7)$$

Constructing an Ising model for CTSP is exactly the same as TSP except for extra allowed visiting sequences, as shown in Fig. 2e. This would lead to a modification of the coupling matrix of $J$ according to Eq. (7) (see the deduction of $J_{CTSP}$ in Supplementary Note 6). From Fig. 2f we can clearly see the differences between $J_{CTSP}$ and $J_{TSP}$. This algorithm of CTSP fits for arbitrary constraints of visiting sequences as well as their combinations.

**Experimental demonstration of 9-city TSP**

We first run a 9-city TSP in the 80 SMTJ-based Ising computer at a relatively low but non-zero effective temperature to examine the intrinsic annealing in SMTJ. The iteration time is set comparable to the longest retention time of SMTJs to avoid reading previous spin states. In our experiments, we set the iteration time as 0.1 ms. As shown in Fig. 3a, as the effective inverse temperature ($c$) is increased quickly to 0.5, the system converges rapidly to a low energy state within 50 iterations and reaches the ground state after 4000 iterations. It should be noted that the intrinsic stochasticity in SMTJs helps the system escape from local minima without an extra annealing process, as shown in the right inset of Fig. 3a. Fig. 3b illustrates the evolution of 9 spins out of 81 spins. The evolution of all 81 spins can be found in Supplementary Note 7. We choose four states in Fig. 3a to inspect the traveling path in 3c and their Ising spins, namely $s_{i,j}$, as shown in Fig. 3d. The yellow square in Fig. 3d represents $s_{i,j} = 1$ (visited) and the blue square represents $s_{i,j} = -1$ (not visited). In an initial state A, the spin states are randomly set and then converge to a relatively low energy at state B. State C is an intermediate solution during the annealing process. State D is the optimal solution satisfying two constraints of the TSP. Because we anneal the system to a relatively low but non-zero temperature so that the convergence to a sub-optimal state could be guaranteed, and at the same time, the intrinsic randomness in SMTJ helps the system to escape from local minima and find a ground state quickly. We test 10 different random initial states each with 5000 iterations and find that in all cases the system can obtain a relatively small energy, as shown in Supplementary Note 8. However, there is a probability that the system jumps out of the ground state because of the non-zero temperature. If we continue to observe the evolution in a large timescale, the system would move back to the global minimum state. In some cases, where the speed and near-optimal solution matter but the accurate optimal solution is not, the number of iterations can be chosen to be small.

Further annealing of the system to a lower effective temperature may guarantee the convergence of the computation. Here we use linear annealing ($c$ increases linearly from 0.2 to 1.8) as an example to examine the convergence of this algorithm in a very large-iteration limit. In Fig. 3e we can find the first global minimum energy appears after 16500 iterations, and converge to the ground state after 40000 iterations. Temperature schedules can be optimized to reduce iteration numbers, e.g. increase the effective temperature in the first few time steps, and then decrease gradually, or learned by the reinforcement learning method[37]. In practice, we use one memory to store the minimum energy state during the computation, and another memory to record the final energy state. We take the minimum value of these two results as the solution. Fig. 3f shows the success probability (defined as finding the optimal path) of TSP with varying node size. The success probability of 9-city TSP reaches 95% after $10^4$ iterations. The success probability with the parameter $w$ in Eq. (5) which determines the relative strength of the constrain term and distance term is also discussed. If the $w$ is too large, then the probabilities of violations, namely the invalid path, would increase, as shown in Supplementary

Note 8. If $w$ is too small, then the effect of the distance term is small, which results in a slower convergence to the ground state.

The advantages of this annealer are threefold: 1) Selective working modes by using different temperature schemes. One is the probabilistic sampling mode working at a constant temperature, which is similar to an asynchronous probabilistic computer[4]; the other is the annealing mode conducted by reducing the effective temperature; 2) Fast speed and low power consumption to find the ground state because of the intrinsic annealing properties in SMTJ. 3) Global annealing outperforms probabilistic sampling in achieving efficient convergence, especially for large-scale problems.

**Compressing 70-city TSP to 80-node Ising computer**

Generally, the number of spins required for an $N$-city TSP is $(N-1)^2$, which limits the scalability of TSP on state-of-the-art computing systems. Here, we propose a graph Ising compressing algorithm based on CTSP that can significantly reduce the number of spins and interactions for solving a TSP. Fig 4a is an example of how we apply this algorithm to our 80-node SMTJ Ising computer for solving a 70-city TSP (4761 nodes, st70 data set from TSPLIB[38]). The major steps of this algorithm can be described as follows: (a) divide the cities into several smaller groups until the number of cities in each group is less than 10 by GP method; (b) solve TSP within each group separately; (c) integrate neighbouring groups to obtain an initial path of the whole group; and (d) optimize the path in (c) by a CTSP window sliding over the whole map.

It is worth mentioning that GP is also an Ising problem. When converting a global TSP into local TSPs, using GP would be more hardware-friendly for our Ising computer compared to other clustering algorithms. It is based on the idea that the original graph can be separated into multiple sub-graphs depending on the Euclidean distance. The number of spins needed for solving GP is ~$N$ and thus, GP is quite efficient for local TSPs since the problem size can be reduced to $\sim (N-1)^2/a$, where $a$ is the number of groups, and each TSP can be optimized independently (see GP mapping in Supplementary Note 9).

The final step (d) is based on CTSP, where a rectangular window slides over the path and cuts it into several disconnected lines, among which the two longest lines are chosen and the edge cities are connected as a circular path (See Supplementary Note 10). The CTSP is solved within each window for sub-area optimization without changing the visiting order of edge cities. After this, the two lines at the edge cities are opened and CTSP is carried out again after sliding to the next window. GP-CTSP-based optimization algorithm provides an efficient way of finding near-optimal solutions for large-scale TSP on limited hardware resources.

Fig. 4b shows the comparison of numbers of spins for different TSPs by a conventional Ising method[9], cluster Ising method[39], and our method. The required number of spins in our method is relatively unchanged for various TSPs, while that of other methods increases substantially with the scale of the problem. Fig. 4c shows the total path of 70-city TSP as a function of iteration number using different SA-based algorithms, including symbiotic organisms search[40], ant colony optimization[41], multi-offspring genetic algorithm[42], and gene-expression programming[7]. Finally, we obtain the near-optimal path with a total energy of 700.71, which is slightly higher than the optimal solution of 675. However, the iteration number for an optimized solution is $4.9 \times 10^6$ by our method, which is two to three orders lower than that of SA-based algorithms running on Intel Core-i7 CPU[7] with the main frequency of 3 GHz, as shown in Fig. 4c.

**SMTJ-MRAM architecture towards large-scale Ising computer**

The above experimental demonstration shows "Ising computer" with 80 SMTJs is capable of finding a near-optimal solution to a medium-scale NP-hard problem. In this section, we propose an SMTJ-MRAM architecture for large-scale Ising computer, which can be integrated by using modern MRAM and CMOS technologies. The core part of SMTJ-MRAM consists of SMTJ bit cells organized as a crossbar array, integrated with row decoders and read sense amplifiers (RSA), as shown in Fig. 5a. Each SMTJ bit cell contains one select transistor and one SMTJ (1T1MTJ), whereas the gate of the select transistor is driven by word lines (WL), and the source of all bit cells are connected to the ground. Each bit line is assigned with an RSA. The current flows through SMTJ can be continuously adjusted by Vin of RSA, and the state of SMTJ can be read by RSA at the same time, showing Vdd or 0 at Vout. Fig. 5b illustrated the circuit of RSA, in which two clamp transistors control the current flow through the bit cell path and reference path by the gate voltage (Vin), and a current mirror is used to guarantee the same current of the above two paths. Then different voltages would show in the Q and QB point when an SMTJ is in the AP or P state. Fig. 5c shows the voltage of the Q and QB changing with different Vin. A large range of linear current region of 1~15 µA can be achieved by the proper design of RSA, and the large window between Q and QB enables for a voltage sense amplifier (VSA) to distinguish the AP and P state of an SMTJ under the control signal of SEN (see details in Supplementary Note 11). Particularly, a voltage equalization circuit (VEC) is designed for initializing VSA to avoid incorrect readout. Fig. 5d shows the signals to control and read bit cells. In phase 0 (PH0), one row of SMTJs is selected by WL, and Vin prepared by peripheral circuit is applied to the corresponding RSA. EQ is set high to initialize Q, QB and Vout as Vdd/2. In phase 1 (PH1), the SMTJ fluctuates from the falling edge to the next rising edge of EQ. Finally, in phase 2 (PH2), RSAs read the data of one row in parallel at the falling edge of SEN. After the first row has been retrieved, the partial sum starts to be computed in FPGA. Meanwhile, the same process for the second row can be started, so and so forth. To avoid reading the previous state, the duration of PH1 is preferred to be comparable with the retention time of SMTJ, which limits the main frequency of the system.

We perform circuit simulation of a 4 Kb SMTJ-MRAM system. The system is divided into 4 PEs, each PE contains 8 columns (3-bit BL) and 128 rows, or 16 columns (4-bit BL) and 64 rows (see details in Supplementary Note 11). We compare the above two designs with individual channel architecture (Fig. 1d). Fig. 5e shows that although the speed of SMTJ-MRAM architecture is slower than individual channel architecture, the energy consumption is greatly saved by using fewer readout circuits (RSAs). Therefore, the energy efficiency of 4 Kb SMTJ-MRAM with the 3-bit BL is 1.9 times higher than that of individual channel architecture. The 4 Kb SMTJ-MRAM with the 4-bit BL also shows a huge improvement of 52083 times higher performance density. From the results, we draw the conclusion that SMTJ-MRAM architecture is more advanced in energy efficiency and performance density, which is preferred for very large-scale Ising computers. The bit of SMTJ-MRAM can be chosen according to various scenarios considering different requirements of time and power.

**Discussion**

We compare our system with other state-of-art Ising solvers, including CMOS annealer (Intel Core i7 processor)[7], quantum annealer (D-Wave 2000Q)[16,17], CIM with FPGA[26], memristor Hopfield neural networks (mem-HNN)[43], and phase-transition nano-oscillators (PTNO) [28] in solving 4761-node TSP70, as shown in Table 1. We use the experimental data for benchmarking from literature, and two kinds of SMTJs for comparison. One is our perpendicular anisotropy SMTJ device and the other is assuming recently reported in-plane anisotropy SMTJ with a retention time of 8 ns[44,45]. The major attributes are the main frequency (defined as 1/iteration time), power, time-to-solution as well as energy efficiency (defined as solutions per second per watt). As quantum computers, CIM, mem-HNN, and PTNO only demonstrated ~100-node max-cut problems, we estimate the time-to-solution for solving TSP70 by assuming that the algorithm and the total number of spins these works should observe to find a near-optimal solution is the same as our work (see details in Supplementary Note 12).

Here, we set 80-spin Ising computer as a standard and fix the number of iterations of 400,000 for a good solution to TSP70. Only Ising computing parts are calculated for power consumption.

In Table 1, although the main frequency of CPU is the highest among all candidates, the time-to-solution and energy efficiency is lower than our SMTJ-based approach. This is due to the redundant logic and data transfer delay between the memory and PEs in a conventional von-Neumann architecture. The SMTJ-based approach currently outperforms the quantum annealer both in power consumption as well as time to solution. The power of quantum annealer is huge which needs to be optimized further for real applications. CIM is another promising architecture with a fast speed and acceptable power consumption. Current coherent Ising machine systems are proof-of-concept systems which are not at present optimized for energy efficiency. Mem-HNN has a relatively fast speed assuming the 180-nm CMOS technology. However, the required number of devices is large, which limits the integrated density. The PTNO approach uses capacitors or resistors to mimic spin coupling, whose main frequency would be limited by the system scale and parasitic effects. It is reported that the ideal main frequency would decrease from 500 to 87 MHz when the system scale increases from 8-node to 100-node[28]. Our SMTJ-based Ising computer outperforms other approaches with low power consumption of 0.64 mW (Supplementary Note 12).

We experimentally demonstrate perpendicular MTJs with a retention time of ~ 0.1 ms and solve TSP70 Ising problems at an energy efficiency of 39.1 solutions per second per watt. Furthermore, we simulate an SMTJ-MRAM Ising computer with 4 Kb SMTJs from our experiments and 40 nm commercial CMOS technology. The energy efficiency for solving TSP70 by using the same algorithm can reach 74 solutions per second per watt. By using reported in-plane SMTJ[44] and advanced CMOS, the system could obtain the highest energy efficiency of $8.3 \times 10^3$, which shows several orders of improvement over other approaches. This result suggests that an SMTJ-based Ising computer can be a good candidate for solving dense Ising problems in a highly energy-efficient and fast way. To solve larger size problems, we can either fabricate more SMTJs on-chip (i.e. the scale-up scheme) or reuse available SMTJs (i.e. the time-multiplexed scheme). Moreover, our system could be highly scalable using commercial MRAM technology.

In summary, we have experimentally demonstrated an intrinsic all-to-all Ising computer based on 80 SMTJs, and solved 9-city TSP with the optimal solution. Furthermore, a compressing strategy based on CTSP and GP was proposed to experimentally solve 4761-node 70-city TSP on an 80-node system with a near-best-known solution as well as ultra-low power consumption. An SMTJ-MRAM architecture is proposed for large-scale Ising computers. Our system provides a feasible solution to fast, energy-efficient, and scalable Ising computing schemes to solve NP-hard problems.

**Methods**

**Sample growth and device fabrication**

Thin film samples of substrate/[W (3)/Ru (10)]$_2$/W (3)/Pt (3)/Co (0.25)/Pt (0.2)/[Co (0.25)/Pt (0.5)]$_5$/Co (0.6)/Ru (0.85)/Co (0.6)/Pt (0.2)/Co (0.3)/Pt (0.2)/Co (0.5)/W (0.3)/CoFeB (0.9)/MgO (1.1)/CoFeB (1.5)/Ta (3)/Ru (7)/Ta (5) were deposited via DC (metallic layers) and RF magnetron (MgO layer) sputtering on the Si substrates with thermal oxide of 300 nm with a base pressure of less than $2 \times 10^8$ Torr at room temperature. The numbers in parentheses are thicknesses in nanometres. To fabricate the superparamagnetic tunnel junctions, bottom electrode structures with 10 μm width were firstly patterned via photolithography and Ar ion milling. MTJ pillar structures of ~50 nm diameter for the superparamagnetic behavior were patterned by using e-beam lithography. The encapsulation layer of Si$_3$N$_4$ was in-situ deposited after ion milling without breaking vacuum by using RF magnetron sputtering, and top electrode structures with 10 μm width were patterned via photolithography and the top electrode layers of Ta (5 nm)/Cu (40 nm) were deposited by using DC magnetron sputtering.

**MTJ characterization by probe station**

The setup includes a source meter (Keithley 2400) for supplying DC bias currents and a data acquisition card (NI-DAQmx USB-6363) for the read operation. A single SMTJ operation cycle comprises two steps (i.e. bias and read). A small DC input current with an amplitude of 1 - 20 µA is applied to SMTJ. Simultaneously, the DAQ card reads the voltage signal across the SMTJ at a maximum sampling rate of 2 MHz. The MTJ switching probability varies in accordance with the applied current amplitude. The retention time of MTJ is determined from random telegraph noise measurements, as shown in Supplementary Note 3.

**Ising PCB**

80 SMTJ arrays and peripheral circuits are integrated on a 12 cm × 15 cm PCB, controlled by an MCU (Arduino Mega 2560 Rev3, as shown in Supplementary Fig 1). Four 12-bit rail-to-rail DACs (AD5381) with 160 output channels in total are used to generate analogue DC inputs for PE and comparator arrays. Half of the DAC output channels are used to provide stimulation to the gate terminal of NMOSs (2N7002DW-G), and others are used to provide reference voltages to comparators (AD8694). The drain voltages of NMOS are compared with reference voltages and generate outputs parallelly. Outputs of comparator arrays are read by MCU through four multiplexers (FST16233) and then are calculated to obtain new inputs for DACs. The supply voltage of the PCB board and SMTJs is 5 V and 0.8 V, respectively. The value of resistors in each computing unit can be designed to adjust the centre of sigmoidal curves, detailed information can be found in Supplementary Note 3.

36. Mitra, D., Romeo, F. & Sangiovanni-Vincentelli, A. Convergence and finite-time behavior of simulated annealing. *Advances in Applied Probability* **18**, 747–771 (1986).

37. Mills, K., Ronagh, P. & Tamblyn, I. Finding the ground state of spin Hamiltonians with reinforcement learning. *Nat Mach Intell* **2**, 509–517 (2020).

38. MP-TESTDATA - The TSPLIB Symmetric Traveling Salesman Problem Instances. http://elib.zib.de/pub/mp-testdata/tsp/tsplib/tsp/index.html.

39. Dan, A., Shimizu, R., Nishikawa, T., Bian, S. & Sato, T. Clustering Approach for Solving Traveling Salesman Problems via Ising Model Based Solver. in *2020 57th ACM/IEEE Design Automation Conference (DAC)* 1–6 (IEEE, 2020). doi:10.1109/DAC18072.2020.9218695.

40. Ezugwu, A. E.-S., Adewumi, A. O. & Frîncu, M. E. Simulated annealing based symbiotic organisms search optimization algorithm for traveling salesman problem. *Expert Systems with Applications* **77**, 189–210 (2017).

41. Mohsen, A. M. Annealing Ant Colony Optimization with Mutation Operator for Solving TSP. *Computational Intelligence and Neuroscience* **2016**, 1–13 (2016).

42. Wang, J., Ersoy, O. K., He, M. & Wang, F. Multi-offspring genetic algorithm and its application to the traveling salesman problem. *Applied Soft Computing* **43**, 415–423 (2016).

43. Cai, F. *et al.* Power-efficient combinatorial optimization using intrinsic noise in memristor Hopfield neural networks. *Nat Electron* **3**, 409–418 (2020).

44. Hayakawa, K. *et al.* Nanosecond Random Telegraph Noise in In-Plane Magnetic Tunnel Junctions. *Phys. Rev. Lett.* **126**, 117202 (2021).

45. Safranski, C. *et al.* Demonstration of Nanosecond Operation in Stochastic Magnetic Tunnel Junctions. *Nano Lett.* **21**, 2040–2045 (2021).
20

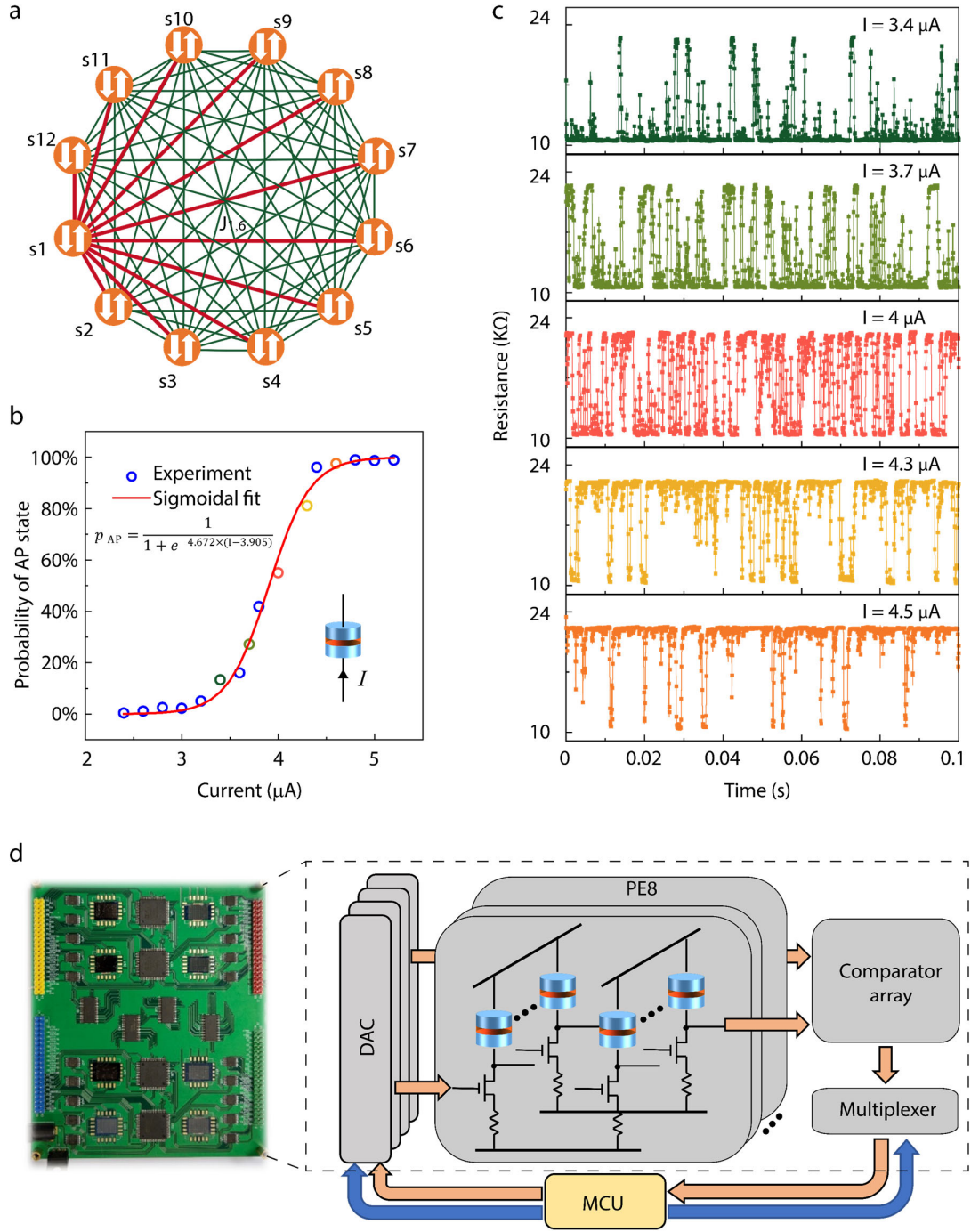

**Fig. 1 Illustration of SMTJ-based Ising computer. a** All-to-all connected 12-spin Ising model. **b** Sigmoidal fit of average resistance of an SMTJ. Insert: diagram of an SMTJ. A tunnelling barrier layer (orange color) is sandwiched by a reference layer and a free layer (blue color). **c** Time-dependent resistance of an SMTJ under different input currents ($I$). **d** Photograph and schematic diagram of SMTJ-based Ising computer. The system contains 8 processing elements (PEs), 4 DACs, a comparator array, and a multiplexer. Each PE has 10 SMTJ computing units. Each computing unit includes a transistor and a resistor to adjust the property into stochastic. Blue lines and orange arrows represent the control and data flow, respectively.

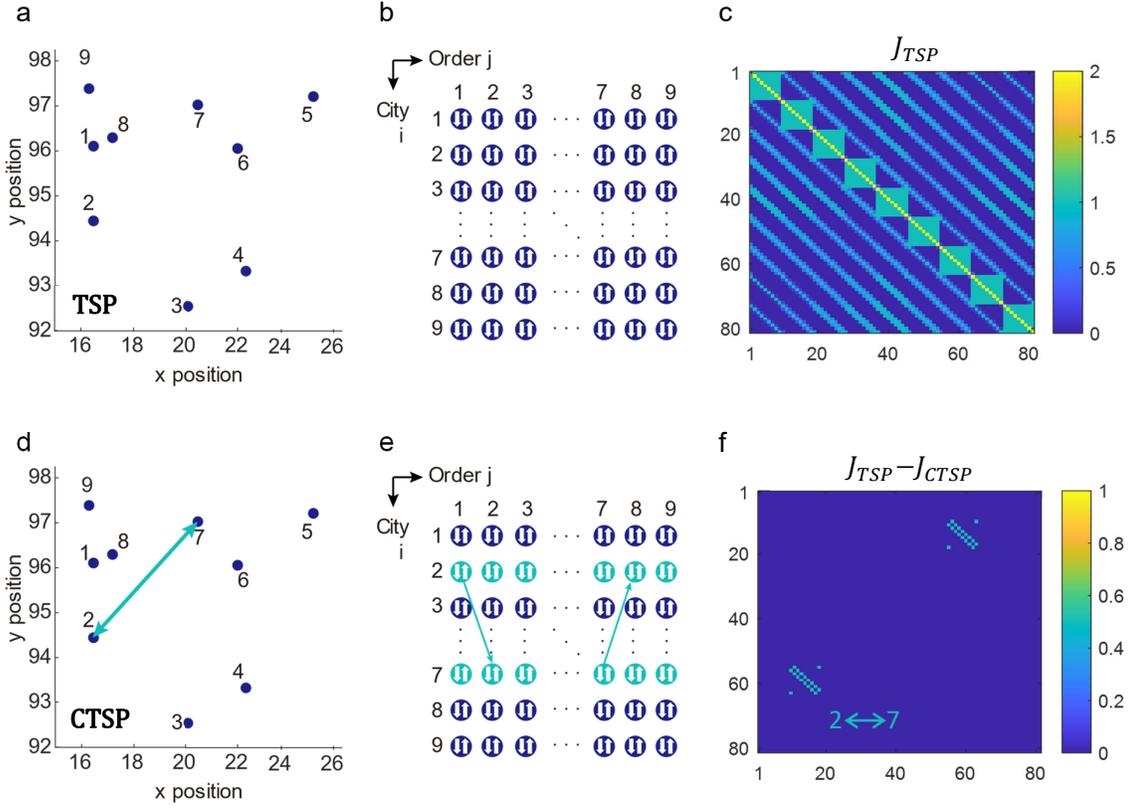

**Fig. 2 Construct Ising model for a 9-city TSP. a** Coordinates of all 9 cities used in this problem. **b** Ising spin representation for 9-city TSP (81 spins). Rows indicate names of cities and columns indicate the visiting order. Each spin can be 1 (visited) or −1 (not visited) in each iteration, depending on the states of other spins, the coupling matrix between each spin and sigmoidal properties with randomness. **c** Colour map of the coupling matrix $J$ of 9-city TSP, and the colour bar represents an effective energy with the unit of $kT$. Here, $k$ is the Boltzmann constant and $T$ is the temperature. **d** CTSP with a fixed vising sequence from city 2 to city 7 or from city 7 to city 2. **e** The Ising spin representation for CTSP which is the same as TSP. Arrows represent allowed vising sequences. **f** Colour map of the difference of coupling matrix $J$ of TSP and CTSP in **d**.

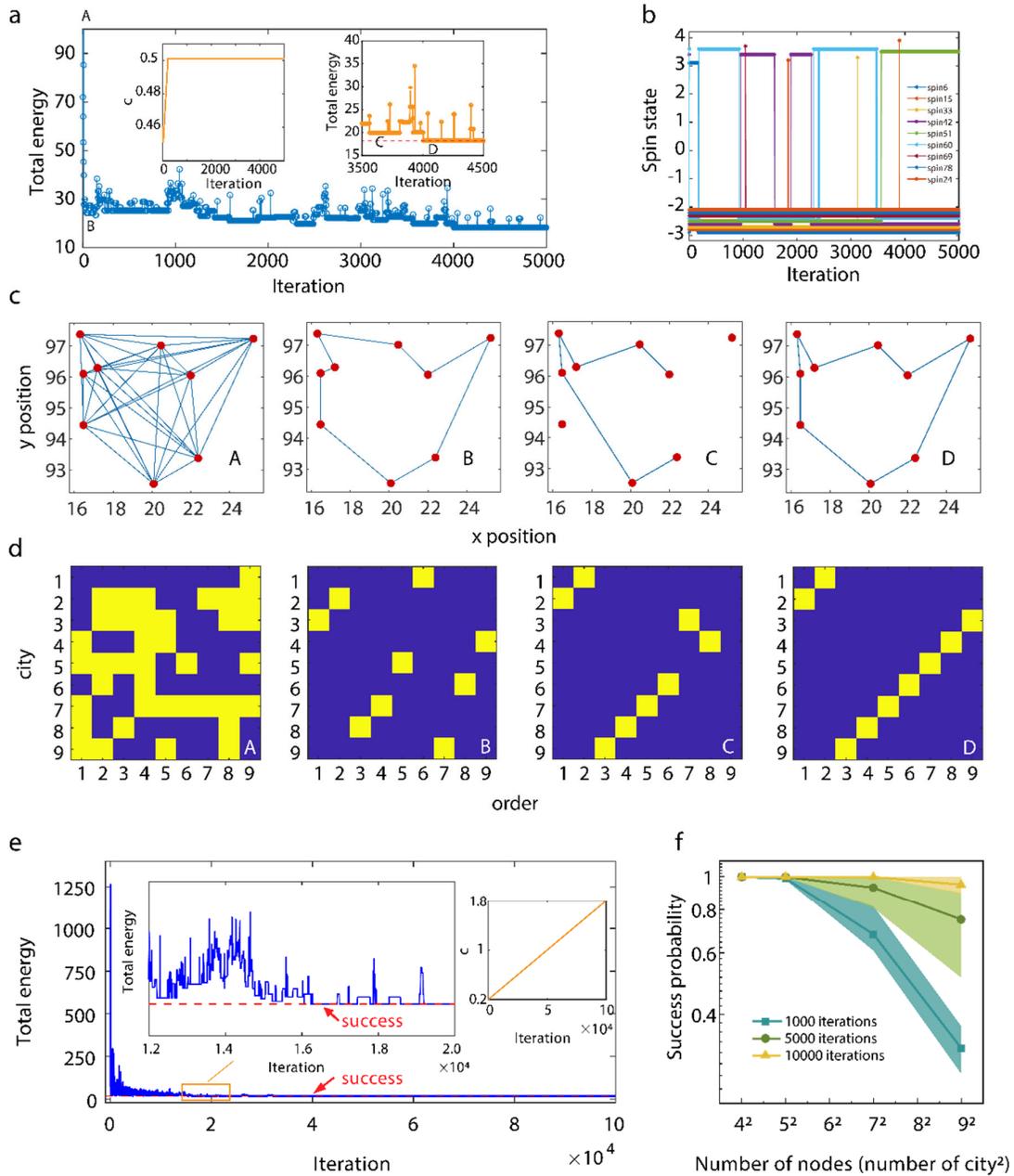

**Fig. 3 Solution to 9-city TSP. a** Total energy transition of 9-city TSP with 5000 iterations (the optimal solution energy of 18.23 corresponds to the dashed horizontal line). Inserts: effective inverse temperature and zoom-in view of 3500-4500. **b** Evolution of 9 representative SMTJ states with 5000 iterations. An offset is used in the y-axis to show each SMTJ clearly. **c** Visiting routes of state A, B, C, and D in **a**. **d** Corresponding Ising spins of state A, B, C, and D in **a**. The yellow squares represent 'visited ($s_{i,j} = 1$)' and the purple squares represent 'not visited ($s_{i,j} = -1$)'. **e** Total energy transition with effective inverse temperature increasing from 0.2 to 1.8. Insert left: zoom-in view of the total energy transition with effective inverse temperature increasing from 0.392 to 0.52. Insert right: effective inverse temperature. The red dashed line represents the optimal path. **f** Success probability of solving TSP of varying the node size.

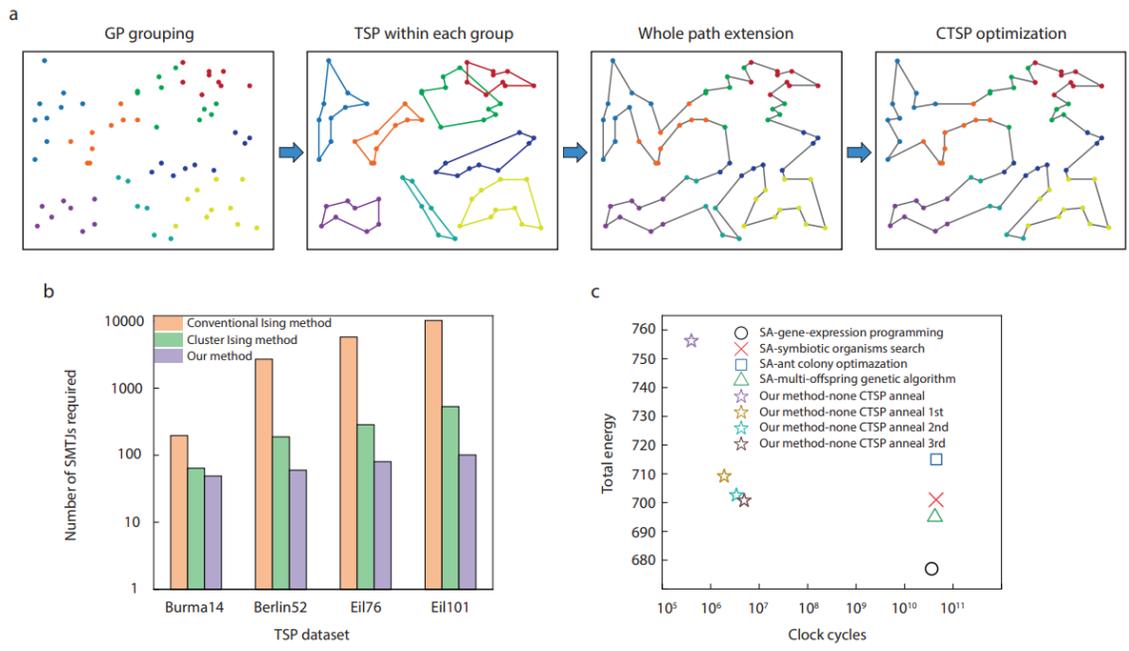

**Fig. 4 Experimental solution of 70-city TSP. a** Optimization algorithm for 70-city TSP. **b** Number of required SMTJs for various problems using different methods. Burma14, berlin52, eil76 and eil101 is TSP of 14, 52, 76, and 101 cities, respectively. **c** Comparison of total energy and time to solutions with different SA-based algorithms, including symbiotic organisms search[40], ant colony optimazation[41], multi-offspring genetic algorithm[42], and gene-expression programming[7]. Our method is tested on our Ising system and others are tested on Intel Core-i7 PC. In this comparison, our system runs at a main frequency of 10 kHz.

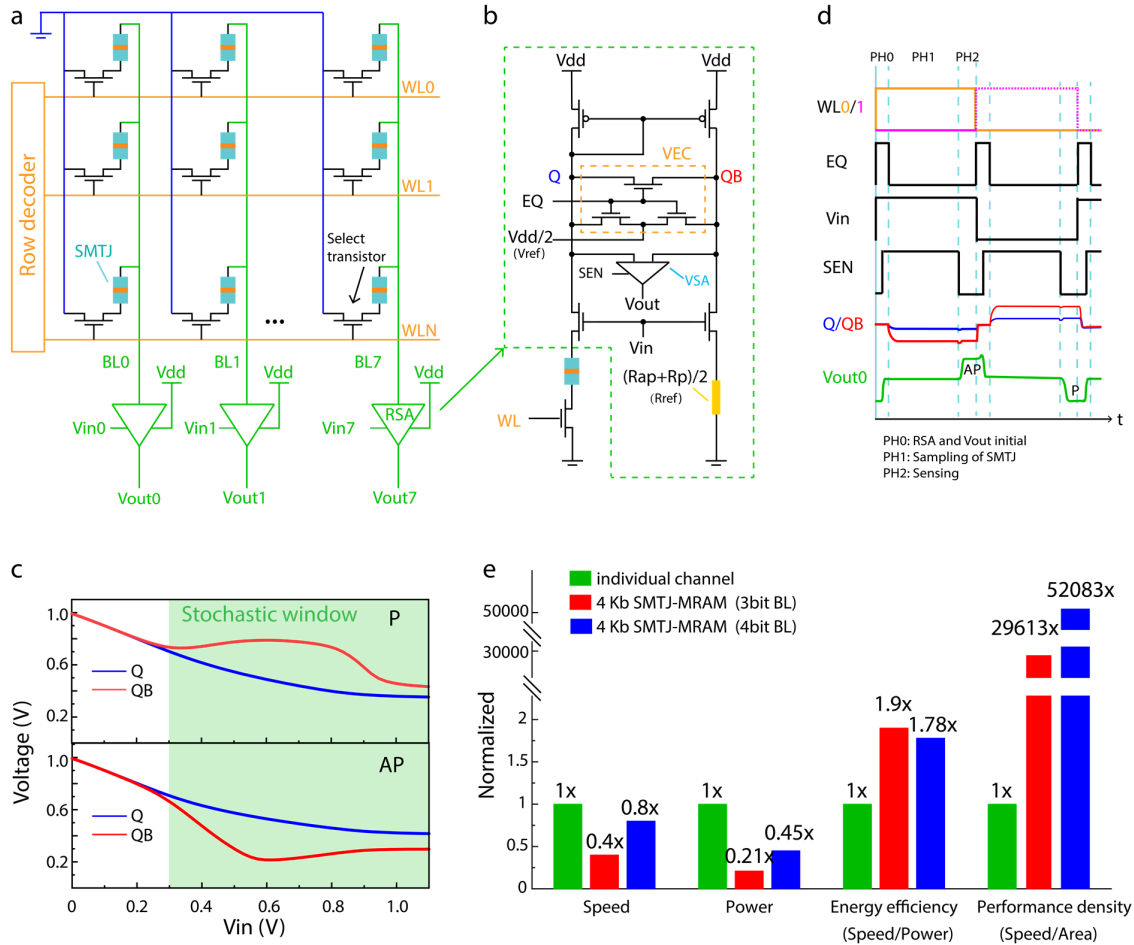

**Fig. 5 SMTJ-MRAM implementation towards large-scale Ising computing. a** SMTJ-MRAM crossbar array. **b** Circuit of one read sense amplifier (RSA), consisting of a current mirror, voltage equalization circuit, sense amplifier and reference resistance. **c** Signals of writing/reading two adjacent SMTJ cells in one BL, selected by WL0 and WL1 in sequence. **d** Voltage of Q and QB when SMTJ at P and AP states showing large stochastic window for Ising computing from 1~15 μA. **e** Simulation results of system-level improvement of 4 Kb SMTJ-MRAM architecture.

## Table 1 Comparison of state-of-art Ising computers

| | CPU | Quantum annealers | CIM | Mem-HNN | PTNO | SMTJ-based Ising computer | | |
|---|---|---|---|---|---|---|---|---|
| **Spin form** | Artificial Ising spin | Qubits | Optical parametric oscillator | Memristor | Nano-oscillators | MTJ | | |
| **Algorithm** | Simulated annealing | Quantum annealing | Coherent computing | Modulate intrinsic noise annealing | Simulated annealing | Intrinsic noise annealing + global annealing | | |
| **Connectivity** | All-to-all | Sparse | All-to-all | All-to-all | All-to-all | All-to-all | | |
| **Room temperature** | Yes | No | Yes | Yes | Yes | Yes | | |
| **Number of nodes solved experimentally (simulated)** | 4761 | 61 | 2000 | 60 | 8 (100) | 4761 | | |
| **Number of devices (bits)** | ~$10^9$ | 2048 | 1 (2048)[d] | 3600 | 8 | 80 | 4096 | |
| **Power** | 65 W | 25 kW | | 120 mW | 2.56 mW[h] | Experiment | Simulation using SMTJ in this work | Simulation using SMTJ in ref 44 |
| | | | | | | 0.64 mW[j] | 0.1344 mW | 0.1344 mW |
| **Size of the computer (chip/system)** | 14 cm$^2$/ 600 cm$^3$ | -/18 × $10^6$ cm$^3$ | > 3× $10^4$ cm$^3$ /- | 64 cm$^2$/1200 cm$^3$ | -/9000 cm$^3$ | 8 cm$^2$/ 900 cm$^3$ | 10752 μm$^2$ /- | < 10752 μm$^2$ /- |
| **Main frequency** | 3 GHz[a] | | 200 kHz[e] | 1 MHz[g] | 301 Hz[i] | 10 kHz | 10 kHz | 1.1 MHz[l] |
| **Time to solution (TSP70)** | 12 s[b] | > $10^4$ s[c] | 80 ms[f] | 320 ms | 1.3×$10^3$ s | 40 s[k] | 100.6 s | 0.9 s |
| **Energy to solution** | 780 J | > 2.5×$10^8$ J | | 3.84×$10^{-2}$ J | 3.4 J | 2.56×$10^{-2}$ J | 1.35×$10^{-2}$ J | 1.2×$10^{-4}$ J |
| **Energy efficiency (1/energy to solution)** | 1.3×$10^{-3}$ | < 4.0×$10^{-9}$ | | < 2.6×$10^1$ | 2.9×$10^{-1}$ | 3.9×$10^1$ | 7.4×$10^1$ | 8.3×$10^3$ |

[a]Estimated average main frequency of Intel Core-i7 PC. [b]Solution time using dataset st70 from ref[7] by using TSP-SAGEP algorithm. [c]Approximation from the result of N=55 node MaxCut problem. [d]1-km fiber with 2048 pulses per second used as Ising spins in a time-multiplexed way. [e]Round-trip time of 5 μs in experiment. [f]Linear approximation from the simulated result of N=2000 node complete graph problem. The time of transferring data from FPGA to computer is 60s×n, where n is the observation time (N=500 in experiments). [g]Main frequency of on-chip computing array using 180-nm CMOS-memristor chip in experiments. [h]Estimated power consumption of main compute kernel. [i]Estimated for 80 node system. [j]Experimentally demonstrated power consumption of main compute kernel. [k]Experimentally tested total solution time using dataset st70 from TSPLIB including computing time, data transfer time, and updating time. [l]Experimentally reported retention time of 8 ns in ref[44] and simulated with 40nm CMOS technology. Unavailable quantities are left blank.